\documentclass{PoS}

\title{The gravitational-wave follow-up program of the Cherenkov Telescope Array}

\ShortTitle{The gravitational-wave follow-up program of the CTA}

\author{\speaker{Monica Seglar-Arroyo$^1$\footnote{E-mail: monica.seglar-arroyo@cea.fr}
}, Elisabetta Bissaldi$^2$, Andrea Bulgarelli$^3$, Alessandro Carosi$^4$, Giancarlo Cella$^5$, Tristano Di Girolamo$^6$, Thomas Gasparetto$^{4,7}$, Giancarlo Ghirlanda$^8$, Brian Humensky$^{9}$, Susumu Inoue$^{10}$, Francesco Longo$^{7}$, Lara Nava$^{11}$, Barbara Patricelli$^{5,12}$, Massimiliano Razzano$^{5}$, Deivid Ribeiro$^{9}$, Fabian Sch\"ussler$^1$, Antonio Stamerra$^{12,13}$, Giulia Stratta$^3$, Susanna Vergani$^{14}$ on behalf of the CTA Consortium\footnote{for collaboration list see PoS(ICRC2019)1177}\\
\llap{$^1$}IRFU, CEA, Universit\'e Paris-Saclay, F-91191 Gif-sur-Yvette, France\\
\llap{$^2$}INFN Sezione di Bari and Politecnico di Bari, via Orabona 4, 70124 Bari, Italy\\
\llap{$^3$}INAF - Istituto di Astrofisica Spaziale e Fisica Cosmica di Bologna, Bologna, Italy\\
\llap{$^4$}LAPP, Univ. Grenoble Alpes, Univ. Savoie Mont Blanc, France\\
\llap{$^5$}University of Pisa and INFN -  Pisa, Pisa, Italy\\
\llap{$^6$}INFN Sezione di Napoli, Via Cintia, ed. G, 80126 Napoli, Italy\\
\llap{$^7$}INFN Sezione di Trieste and Universit\'a degli Studi di Trieste, Trieste, Italy\\
\llap{$^8$}INAF, Osservatorio Astronomico di Brera, Merate, Italy\\
\llap{$^{9}$}Columbia University, New York, USA\\
\llap{$^{10}$}RIKEN, Saitama, Japan \\
\llap{$^{11}$}INAF - Osservatorio Astronomico di Brera, Milano, Italy\\
\llap{$^{12}$}INAF - Osservatorio Astronomico di Roma, Roma, Italy\\
\llap{$^{13}$}Agenzia Spaziale Italiana (ASI), Roma, Italy\\
\llap{$^{14}$}Observatoire de Paris, CNRS, Meudon, France}

\abstract{The birth of gravitational-wave / electromagnetic astronomy was heralded by the joint observation of gravitational waves (GWs) from a binary neutron star (BNS) merger by Advanced LIGO and Advanced Virgo, GW170817, and of gamma-rays from the short gamma-ray burst GRB170817A by the \textit{Fermi} Gamma-ray Burst Monitor (GBM) and INTEGRAL. This detection provided the first direct evidence that at least a fraction of BNSs are progenitors of short GRBs. GRBs are now also known to emit very-high-energy (VHE, > 100 GeV) photons as has been shown by recent independent detections of the GRBs 1901114C and 180720B by the ground-based gamma-ray detectors MAGIC and H.E.S.S. In the next years, the Cherenkov Telescope Array (CTA) will boost the searches for VHE counterparts thanks to its unprecedented sensitivity, rapid response and capability to monitor large sky areas via survey-mode operation. In this contribution, we present the CTA program of observations following the detection of GW events. We discuss various follow-up strategies and links to multi-wavelength and multi-messenger observations. Finally we outline the capabilities and prospects of detecting VHE emission from GW counterparts.}

\FullConference{36th International Cosmic Ray Conference -ICRC2019-\\
		July 24th - August 1st, 2019\\
		Madison, WI, U.S.A.}

\begin{document}

\section{Introduction}

In 2017, the joint observation of gravitational waves (GWs) from a binary neutron star (BNS) merger, GW170817 \cite{gw170817} by Advanced LIGO \cite{LIGO} and Advanced Virgo \cite{Virgo}  and the short Gamma-Ray Burst GRB170817A observed by \textit{Fermi}-GBM and INTEGRAL \cite{grb170817}, marked the beginning of a new era in astronomy. These observations launched an unprecedented multi-messenger follow-up campaign \cite{MMgw170817} which enabled the localization of the optical/infrared counterpart, hosted in the galaxy NGC4993. The subsequent detection and monitoring of the remnant X-ray and radio emission enabled the study of the ongoing processes and set constraints on emission models  \cite{alexandre} \cite{troja} \cite{Ghirlanda}.

For years, various progenitors and potential production mechanisms, such as the mergers of neutron star binaries \cite{blinn}, neutron star-black hole binaries \cite{nakar} and the core collapse of massive stars \cite{woos},  have been proposed to account for the observed GRB emissions. The event GW170817/ GRB170817A provided the first direct evidence that at least a fraction of BNSs are progenitors of short GRBs. GRBs are now also known to emit very-high-energy (VHE, > 100 GeV) photons as has been shown by recent independent detections of GRBs 190114C and 180720B by the ground-based gamma-ray detectors MAGIC \cite{MirzoyanGRB} and H.E.S.S \cite{Edna}.

In the coming years, the next generation imaging atmospheric Cherenkov telescopes (IACTs), the Cherenkov Telescope Array (CTA) will play a crucial role in gamma-ray astronomy due to its unprecedented sensitivity, an order of magnitude better than current instruments. CTA has been designed to be a complementary two-site observatory which focuses on slightly different science topics (which translates to different combinations of telescopes per site) covering an energy range from 20 GeV to 300 TeV.
While the design of CTA-North includes a total of 4 Large Size Telescopes (LSTs) and 15 Medium Size Telescopes (MSTs),the number of telescopes in CTA-South, which is much larger in extension, has been chosen so that it will include three classes of telescope 70 Small Size Telescopes (SSTs), 25 MSTs and 4 LSTs \cite{CTAconsortium}. 
In particular, searches for VHE counterparts in gravitational wave follow-up will benefit from the rapid response and fast slewing capabilities, the low energy threshold of observation and the capability to monitor large sky areas via survey-mode operation. In the scheme of the CTA Key Science Project on transients \cite{CTAconsortium}, gravitational wave transients are ranked as one of the highest priority to be studied. Hence, follow-up strategies are being discussed in order to put in place a competitive, rapid response to alerts, which was the issue of previous studies \cite{Bartos2014,patricelli2016, Bartos2018,patricelli2018}. In addition, the complementarity provided by the two sites of CTA: CTA-North and CTA-South will enable complex, parallelized strategies. 
In this contribution, we briefly describe previous follow-up strategies in IACTs and we present the GW follow-up program proposed for CTA. This program includes the follow-up observation scheduling and the Real-Time Analysis \cite{RTA}. In order to prove the proposed strategy, phenomenological simulations connecting the emission of GW and gamma rays, have been obtained. The GW-EM simulation bank will be described in detail. Then, the gravitational-wave follow up and the subsequent electromagnetic counterpart search are simulated. In the last section, a description of the method to estimate the GW-CTA joint detection rates are given.
\section{Searches of EM counterparts in GW follow-up observations with CTA}
\label{GWfollowupCTA}
\begin{figure}
\centering
 \includegraphics[width=10.0 cm]{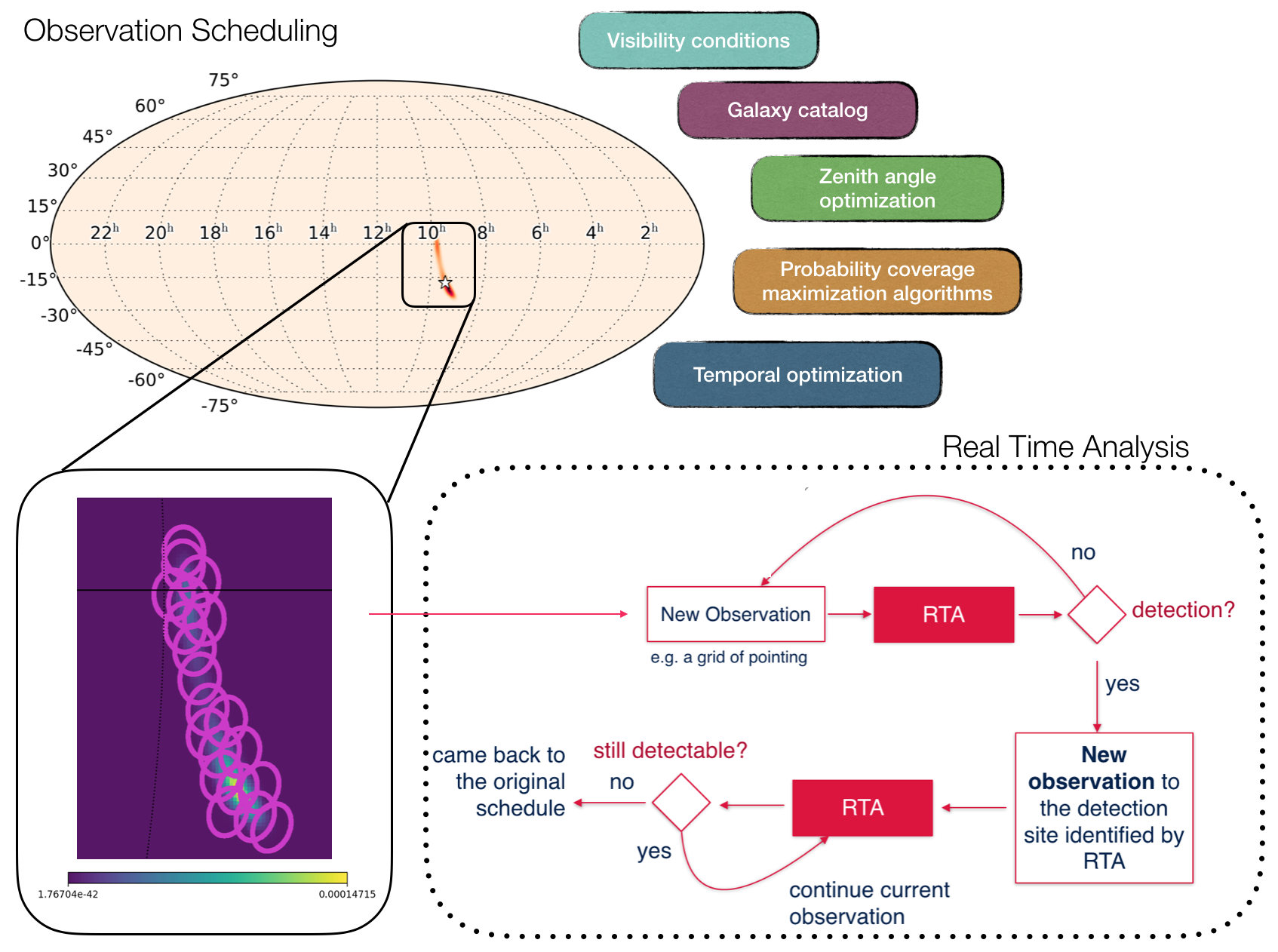}
\caption{Illustration of the gravitational waves follow-up program of CTA}\label{Program}
\end{figure}

In this contribution we present the low-latency gravitational waves follow-up program of CTA. The main two parts of the program, the follow-up observation scheduling and the Real-Time Analysis are shown in Figure \ref{Program}.

\textbf{GW-follow-up observation scheduling}\\

After the reception of the GW alert and sky localisation on both sites, the short-term scheduler determines the visibility window and has the goal of computing the most favorable sky coordinates for the observation, taking into account the array status and observing conditions (weather and night sky background). In particular, the observation scheduling is to optimise the observation in three axes: energy, probability coverage and time, whose motivation and implementation is described in the following:

\begin{itemize}
\item Low-energy coverage. Due to the absorption of shower light during its passage through the atmosphere, which is larger for larger zenith angles, the energy threshold of an observation performed by IACTs depends on the zenith angle under which the source is observed. Based on the soft spectrum of sGRBs observed by \textit{Fermi}-Large Area Telescope \cite{ajello2019}, we prioritize the low-energy domain. To this aim, the scheduling algorithms include a module in which low-zenith-angle observations are favored.

\item Probability coverage maximization. Several techniques have been developed to guide the follow up of gravitational wave events in a smart, efficient way, covering the most probable regions as fast as possible \cite{abadie}. A sequential order of the observations based on the covered probabilities in the 2D-GW localization maps, from the highest to the lowest, enhances the likelihood of covering the EM counterpart in a shorter period of time.
Moreover, the searched region can be reduced, and the chances of detecting the EM counterpart can be increased, by convolving the 3D-localization region of a gravitational wave with a distribution of galaxies which could plausibly host such cataclysmic events \cite{goingthedistance}. These two approaches are considered in scheduling follow-up observations with CTA and used depending on the characteristics of the event, e.g. distance and overlapping with the \textit{avoidance zone}, which is defined by the lack of objects in galaxy catalogs due to the observational bias due to the Galactic Plane. 

\item Dynamic windows. If some of the source parameters are known, such as the distance of the event or the spectral model, or a hypothesis on those is made from modelling, the duration of the observation windows can be estimated \cite{patricelli2018}. In this way, we consider the source parameters related to the EM emission as the isotropic energy emitted by the source E$_{iso}$, the spectral shape, and the temporal evolution. For each observation window the duration T$_{obs}$ is set to the time required to make a 5 sigma detection at the sensitivity that has been quoted for those observation conditions using the instrument response functions (IRFs) \cite{IRF} at a given time post-merger. T$_{obs}$ is selected so that the following condition is fulfilled: 

\begin{equation}
\int_{t_0}^{t_0+T_{obs}} \frac{dF(t)}{dt} dt = F^{int}_{5\sigma}(t_0,t_0+T_{obs})
\end{equation}

This gives us a set of observing times T$_{obs}$ which become larger as the t-t$_{merger}$ increases, since the light-curve evolution of the source decays with time, until the moment when T$_{obs}$ $\rightarrow \inf$ and a 5 $\sigma$ detection is no longer possible. 
\end{itemize}

\textbf{Real-Time Analysis}\\

The Real-Time Analysis (RTA) science alert system is a crucial part of the gravitational wave follow-up program. Each observation is analysed in real time by the RTA pipeline, which is able to detect sub-minute emission,  trigger deeper observations on the region to asses the detection of the potential EM counterpart and issue science alerts at low latencies below 30 seconds to external observatories \cite{Fioretti}. Details of the operation and observations with CTA for different science cases
were defined in the Top Level Use Cases \cite{TopLevelUseCases}. The process in the case of the GW-follow-ups is illustrated in Figure \ref{Program}.

\section{Simulations of VHE counterpart searches in GW follow-ups with CTA}
\label{Simu}
In order to asses the proposed strategy for the GW-follow up program with CTA and with the goal of deriving GW-EM detection rates with CTA at VHE, a set of simulated GW-EM events have been obtained, which are described in the following.

\subsection{Simulations of BNS mergers and GW detections}
\label{GW}
In this work we used the sample of simulated BNS mergers and their GW detection available in the public database GWCOSMoS \cite{GWcosmos}, that is based on the work by Patricelli et al. 2016, 2018 \cite{patricelli2016,patricelli2018}. This database consists of a realistic ensemble of BNS merging systems evenly distributed in the Local universe according to a merger rate of 830 Gpc$^{-3}$ yr$^{-1}$ (see  \cite{patricelli2016,patricelli2018} and references therein), that is within the range estimated after the detection of GW170817 \cite{patricelli2018,gw170817}. The maximum distance considered is  500 Mpc, that is consistent with the expected BNS horizon\footnote{The horizon is the maximum distance at which the interferometers can detect an optimally located, optimally oriented BNS merger.} of Advanced LIGO and Advanced Virgo in their final configuration \cite{prospects}. The GW emission associated with the BNS mergers has been simulated using the ``TaylorT4'' waveforms (see, e.g., \cite{buonanno}), that are constructed using post-Newtonian models accurate to the 3.5 order in phase and 1.5 order in amplitude. The simulated GW signals have been convolved with the GW detector responses, using the sensitivity curves of Advanced LIGO and Advanced Virgo in their final design configuration \cite{prospects} and the data obtained in this way were  analyzed with the matched filtering technique \cite{wainstein,dalcanton,veitch,adams,usman,cannon,messick,nitz}. An 80\% independent duty cycle has been assumed for each interferometer over a 1-year run of data taking, 
and the signals were considered as GW candidates if they were detected by at least two detectors with a time delay between them consistent with the propagation of GWs, with a combined signal-to-noise ratio above 12  \cite{patricelli2016}. 
Finally, for each simulated GW  candidate the GW sky localization has been estimated with BAYESTAR, that is a rapid Bayesian position reconstruction code which computes source location using the output from the GW detection pipelines \cite{bayestar}.

\subsection{Simulation of HE-VHE emission from short GRBs}
\label{GRB}
We associate to each simulated BNS merger a very high-energy emission, simulated under the following conditions. We consider a purely phenomenological approach based on observations of GRBs at GeV energies, mainly by {\it Fermi}-LAT. 
The distance and angle with respect to the line of sight are known for each event in the mock catalog of BNS mergers.
To model the very high-energy emission, we first associate to each event an isotropic-equivalent prompt energy $E_{iso}$.
This is randomly extracted from the intrinsic distribution inferred for short GRBs in \cite{Ghirlanda16}. 
The 0.1-10\,GeV luminosity as a function of time is derived based on the typical properties of LAT GRBs, and in particular of the short event GRB\,090510. 
During the initial phase of the afterglow emission (before deceleration) the flux is assumed to be  proportional to $t^2$ (as expected for homogeneous medium); the afterglow onset is fixed at $t_{peak} = 3$\,s; during the deceleration phase the luminosity decreases as $t^{-1.4}$. To normalize the light curve, we use the correlation found in \cite{Nava14} between E$_{iso}$ and L$_{\textrm{LAT}}^{\textrm{t=}60\textrm{s}}$, for a sample of 10 LAT GRBs, including also GRB\,090510.
For the spectral shape we consider a simple power law with photon index -2.1 ($N_E \propto E^{-2.1}$) and normalization derived from the integrated luminosity 0.1-10\,GeV. The spectrum is extrapolated up to 10\,TeV.
The light curves and spectra generated using this method refer to emission detected on-axis. We then consider the viewing angle $\theta_{view}$ and apply a correction assuming a homogeneous jet, with jet opening angle of 5 degrees, following the prescription given in \cite{Granot02}.

\subsection{Simulations of Observation Scheduling of GW follow up with CTA using \texttt{Gammapy}}
The algorithm described in Section \ref{GWfollowupCTA} has been used to derive an observation schedule of a simulated gravitational wave from the GWCOSMoS catalog. 

\begin{itemize}
\item \textbf{Alert injection and GW follow-up observation scheduling}. The gravitational wave is injected in the gravitational-wave follow-up pipeline at a random time. The set of observation windows are obtained by considering the latency time for the alert reception T$_{alert}$ which is set to $\simeq$ 3 minutes, the initial slewing time of the telescopes T$_{slew}$, estimated to be T$_{slew}$ $\simeq$ 30 seconds and the observation times derived as explained in Section \ref{GWfollowupCTA}. Each observation window T$_j$ is given by T$_j$= T$_{alert}$ +  T$_{slew}$ + $\sum^{j-1}_{1}$T$^j_{obs}$. 

\item \textbf{CTA observation searching for an EM counterpart}.
The GRB emission is simulated and analysed using the open-source Python package for gamma-ray astronomy Gammapy \cite{deil}. The performance of the CTA instrument is described by the IRFs obtained from detailed Monte Carlo simulations \cite{IRF}. The IRFs include information about the effective area, the point spread function, the energy dispersion and the background, and are quoted for a given site, zenith angle, night-sky background and observation time. The IRFs are used in order to extract the background and exposure maps for a given livetime and pointing coordinates. A 4D skymodel for the source is evaluated considering the time evolution, spatial and spectral properties which are given by the GWCOSMoS and the GRB simulations (Section \ref{GW} and \ref{GRB}). The 4D skymodel is convoluted with the IRFs, returning a map with the predicted number of counts. A set of simulated CTA observations for the GW follow up is shown in Figure \ref{SimulationPointing}. 

\begin{figure}
\centering
 \includegraphics[width=7.5 cm]{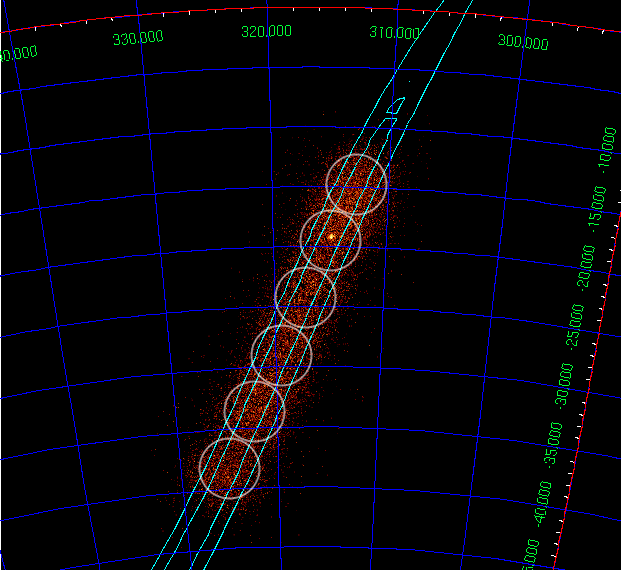} 
\caption{Simulated counts for a gravitational-wave follow-up where 6 observations of 15 seconds have been taken assuming a conservative FoV=2.5$^{\circ}$ (white) which corresponds to the $FoV_{LST}$ in the CTA array design. Cyan contours illustrate the localization uncertainty of the injected gravitational wave.}\label{SimulationPointing}
\end{figure}

\item \textbf{Analysis of the CTA scheduled observations}. The observations are analyzed on a run-by-run basis, in order to mimic a real-time response. The analysis is based on a test-statistic (TS) technique, which consists in a single parameter amplitude fit, which finds the roots of the derivatives of the fit statistics using root finding algorithms, following \cite{Stewart}. A TS analysis example of a CTA observation considering a field of view (FoV), FoV=2.5$^{\circ}$, corresponding to the LST FoV in the CTA array design when acceptance is at 50\% is shown in Figure \ref{SimulationObservation}. Note that this consideration is conservative as larger FoV telescopes, i.e. MSTs, are expected to be part of the gravitational-wave follow up.

\begin{figure}
\centering
 \includegraphics[width=7.65 cm]{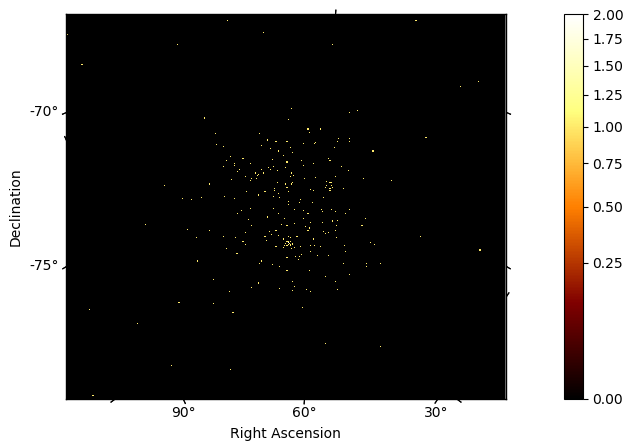} \includegraphics[width=7.35 cm]{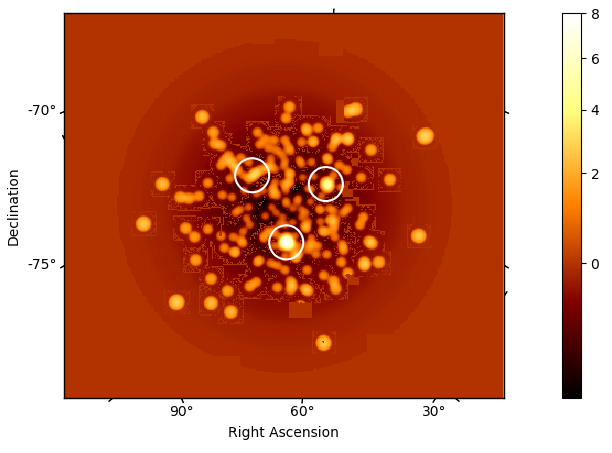} 
\caption{(Left) Simulated counts for a GW follow-up observation which contains the associated simulated GRB source, with T$_{obs}$=2 s (Right) test-statistic analysis of the simulated follow-up observation containing the GRB source. White circles represent the \textit{hot spots} found above 3 sigma, and the GRB is detected at 8$\sigma$. Both figures have been obtained using Gammapy.}\label{SimulationObservation}
\end{figure}

\end{itemize}

\section{Outlook}

The next step of this work is the derivation of the GW-EM detection rates with CTA and the study of the influence of the physical parameters of the source on the derived rates. In this contribution, the simulations which will be used to develop such studies in a future work have been introduced.
By the time CTA will be able to produce science results, the sensitivity of gravitational-wave interferometers will reach design expectations \cite{prospects} and further detectors may have successfully joined the network, such as KAGRA, and LIGO-India at later times.

\section{Acknowledgements}
This work was conducted in the context of the CTA Transients Working Group \cite{TransientsICRC}. We gratefully acknowledge financial support from the agencies and organizations listed here: \href{http://www.cta-observatory.org/consortium_acknowledgments}{http://www.cta-observatory.org/consortium\_acknowledgments}.

\end{document}